\documentclass[11pt]{article}
\usepackage{moriond,epsfig}
\usepackage{graphicx}
\usepackage{amsmath}
\usepackage{enumerate}
\usepackage{indentfirst}
\usepackage{xspace}

\def\Journal#1#2#3#4{{#1} {\bf #2} (#3) #4}

\def\NIMA{{\em Nucl. Instrum. Methods} A}
\def\NPB{{\em Nucl. Phys.} B}

\def\PRL{\em Phys. Rev. Lett.}
\def\PRD{{\em Phys. Rev.} D}
\def\PRC{{\em Phys. Rev.} C}

\def\EPJC{{\em Eur. Phys. J.} C}
\def\AstroP{{\em Astropart. Phys.}}
\def\jinst{{\em J. Inst.}}

\def\be{\begin{equation}}
\def\ee{\end{equation}}
\def\bea{\begin{eqnarray}}
\def\eea{\end{eqnarray}}

\newcommand{\piplus}{\ensuremath{\pi^{+}}\xspace}
\newcommand{\pip}{\ensuremath{\pi^{+}}\xspace}
\newcommand{\piminus}{\ensuremath{\pi^{-}}\xspace}
\newcommand{\pim}{\ensuremath{\pi^{-}}\xspace}

\newcommand{\nn}{\ensuremath{\mbox{N}_{2}}\xspace}
\newcommand{\oo}{\ensuremath{\mbox{O}_{2}}\xspace}

\newcommand{\GeVc}{\ensuremath{\mbox{GeV}/c}\xspace}
\newcommand{\MeVc}{\ensuremath{\mbox{MeV}/c}\xspace}

\newcommand{\rad}{\ensuremath{\mbox{rad}}\xspace}

\begin{document}
\vspace*{1cm}
\title{
  HARP COLLABORATION RESULTS ON THE PROTON-NUCLEI INTERACTIONS AT FEW GEV
               ENERGIES~\footnote{Talk presented at the XLIII
Rencontres de Moriond on  Electroweak Interactions and Unified
Theories,
La Thiule, 1-8 March 2008}
}
\vspace*{-0.35cm}
\author{ Roumen Tsenov}
\address{Department of Atomic Physics, Faculty of Physics,\\
 St. Kliment Ohridski University
  of Sofia, Sofia, Bulgaria\\ {\bf (on behalf of the HARP Collaboration)} }

\maketitle\abstracts{
Recent results obtained by the HARP collaboration on the measurements
of the double-differential production cross-section of positive and
negative pions in proton interactions with nuclear targets from
Beryllium to Lead are presented. They cover production at small angles
(30-210 mrad) and relatively large momenta up to 8 GeV/c as well as
large angles (0.35 - 2.15 rad) and small momenta (0.1 - 0.8 GeV/c).
These results are relevant for a detailed understanding of
neutrino fluxes in accelerator neutrino experiments, better prediction
of atmospheric neutrino fluxes, optimization of a future neutrino
factory design and for improvement of hadronic generators widely used
by the HEP community in the simulation of hadronic interactions.}

\vspace*{-0.4cm}
%
\begin{figure}
  \begin{center}
    \begin{minipage}[c]{0.58\linewidth}
      \epsfig{file=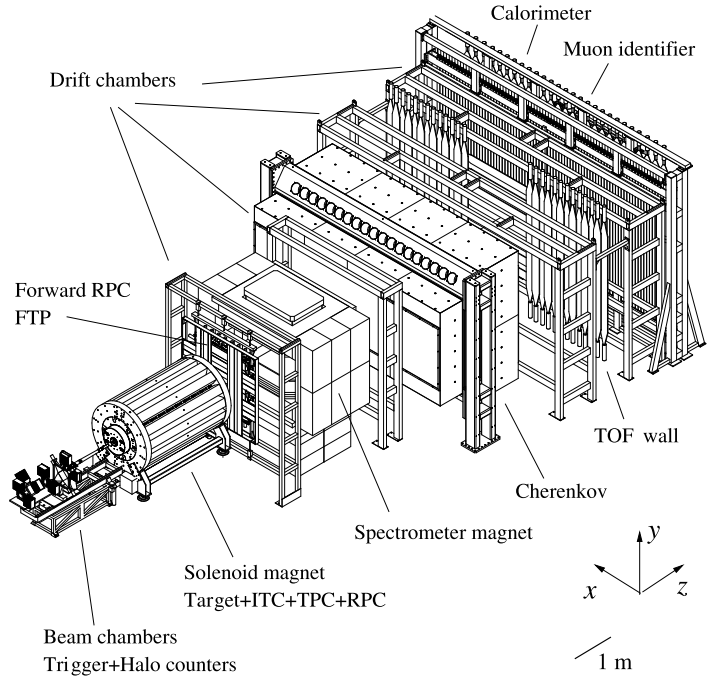, width=\linewidth}
    \end{minipage}\hfill
    \begin{minipage}[c]{0.30\linewidth}
      \caption{Overall mechanical layout of the HARP detector. 
	The different sub-detectors
	are shown. The target is inserted inside the TPC.}
      \label{GeneralLayout}
    \end{minipage}
  \end{center}
\end{figure}

\vspace*{-0.35cm}
\section{The HARP experiment}
\vspace*{-0.20cm}

The HARP experiment~\cite{ref:harp-prop,ref:harpTech} at the CERN PS
was designed to make measurements of hadron yields from a large range
of nuclear targets and for incident particle momenta from 1.5~\GeVc to 15~\GeVc.
The main motivations are the measurement of pion yields for a quantitative
design of the proton driver of a future neutrino factory, a
substantial improvement in the calculation of the atmospheric neutrino
flux
and the measurement of particle yields as input for the flux
calculation of accelerator neutrino experiments,
such as K2K~\cite{ref:k2k,ref:k2kfinal},
MiniBooNE~\cite{ref:miniboone} and SciBooNE~\cite{ref:sciboone}.


\begin{figure}[tbph]
\centering
\includegraphics[width =0.49\textwidth, height=11.0cm]
{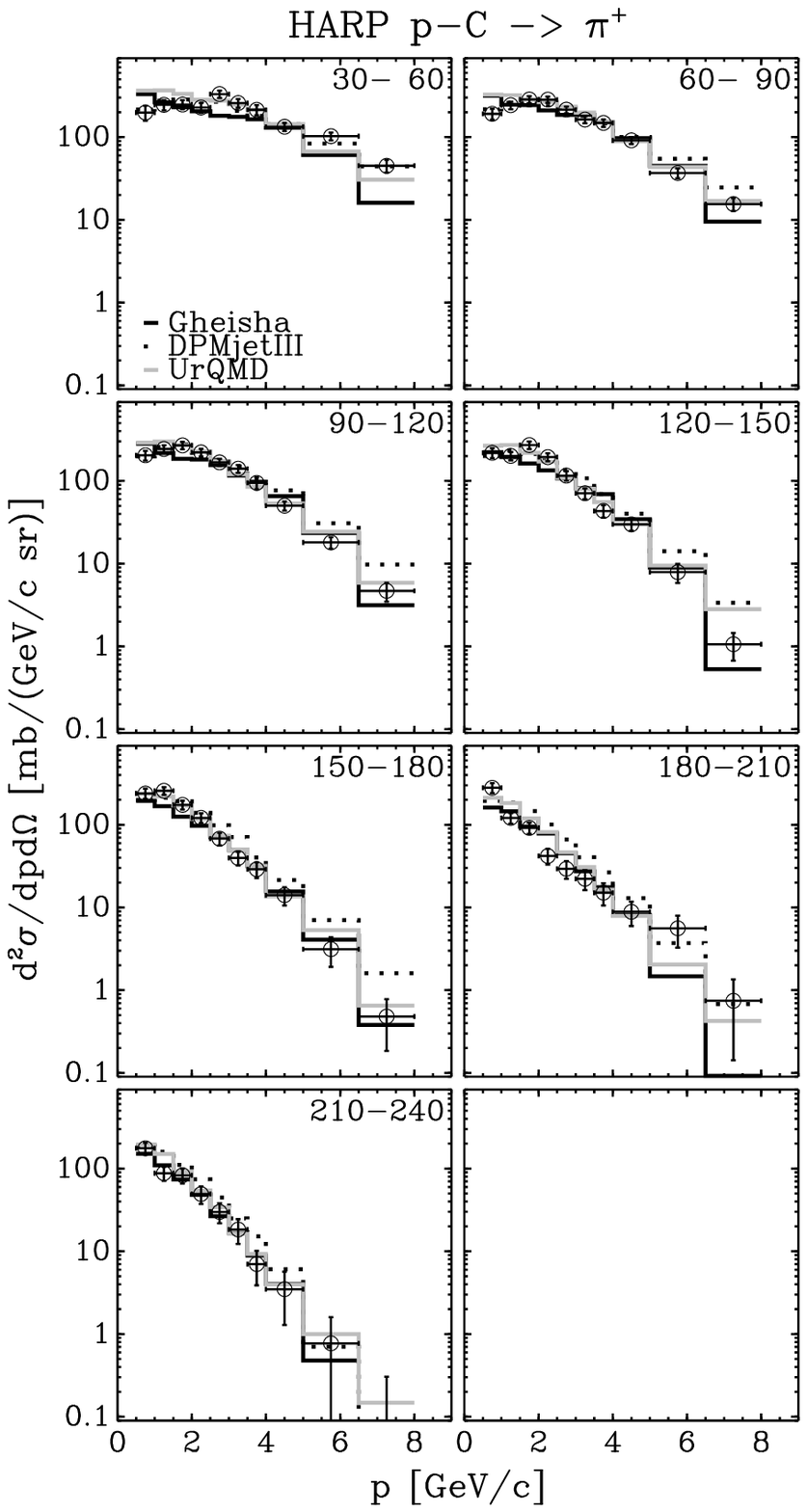}
\includegraphics[width =0.49\textwidth, height=11.0cm]
{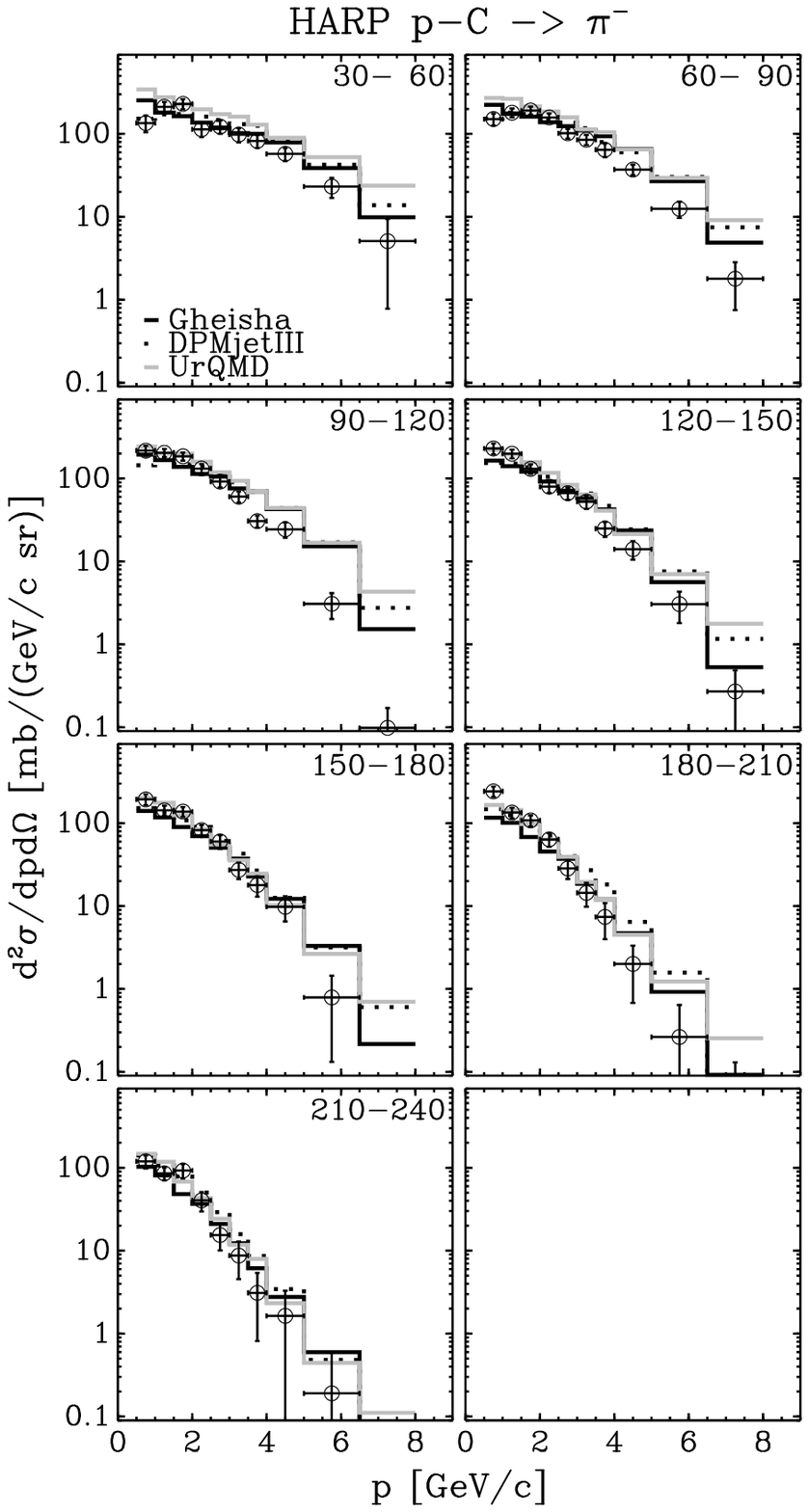}
\caption{Double-differential production cross-section of 
\piplus and \piminus
in p--C reactions at 12~\GeVc (points with error bars)
and comparison with  model predictions.
}
\label{pCpipmModelslog}
\end{figure}

\begin{figure}[tb]
\centering
\includegraphics[width=.49\textwidth]{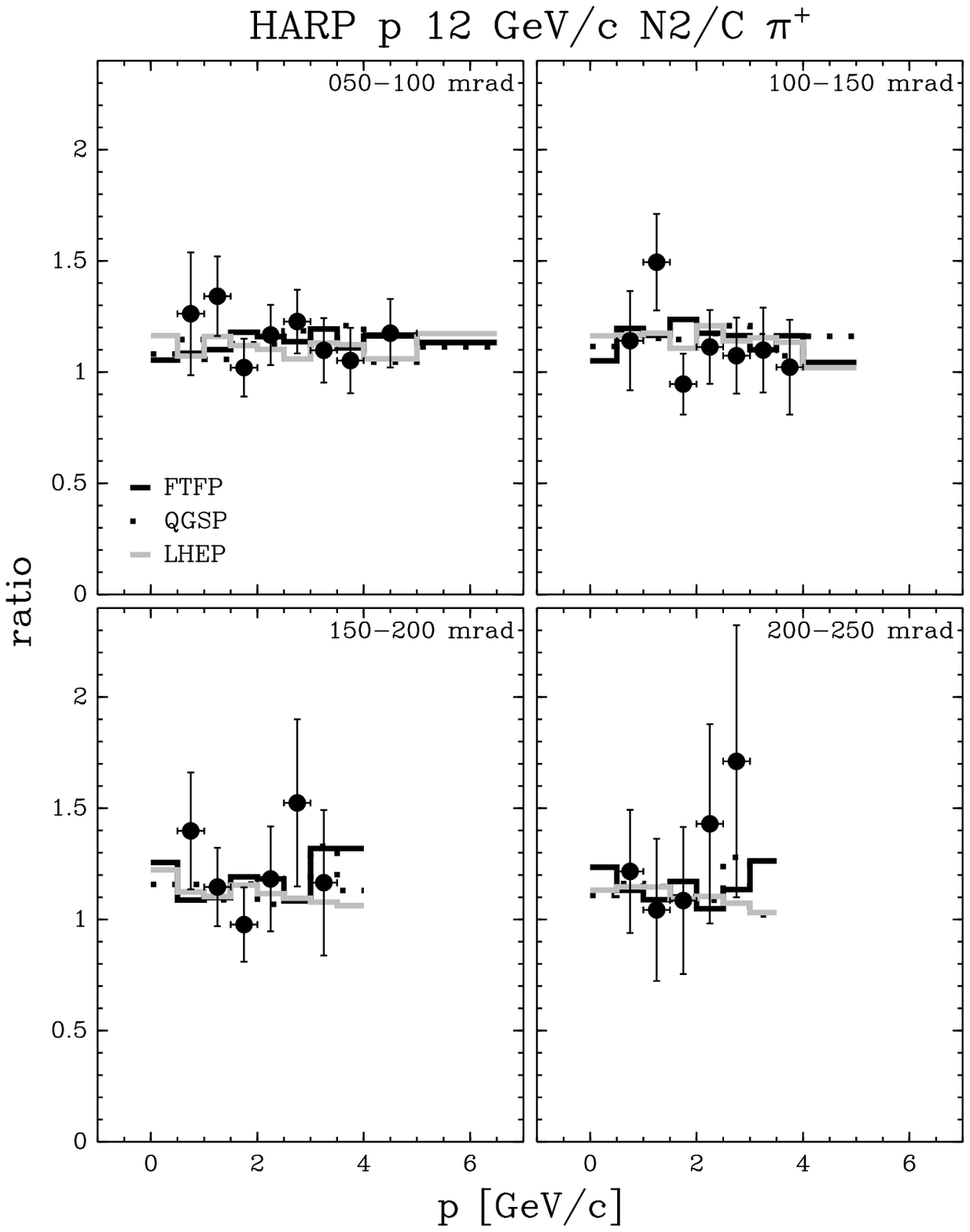}
\includegraphics[width=.49\textwidth]{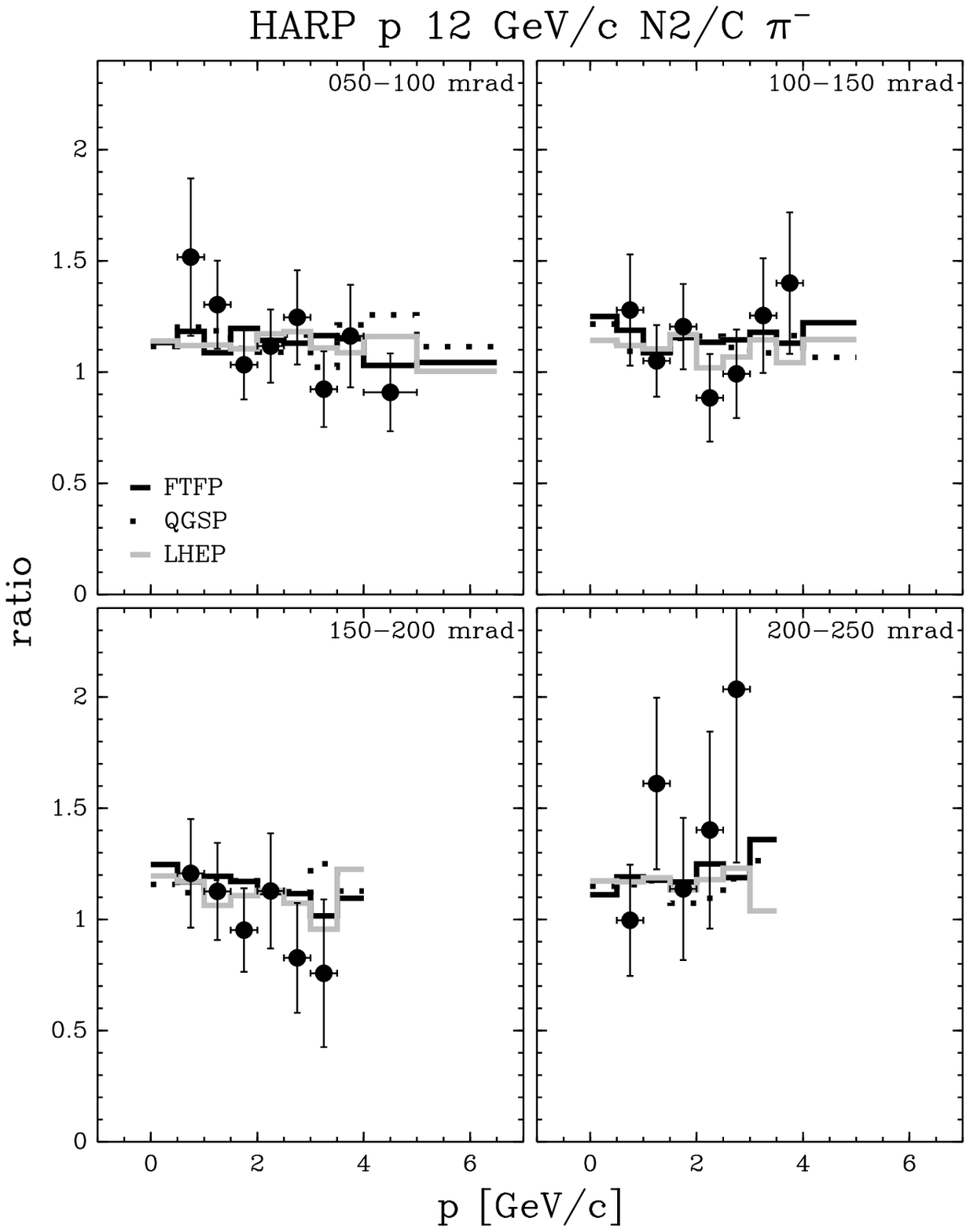}
\caption{
  p--\nn to p--C production
  ratio for \piplus and \piminus
  at 12~\GeVc, compared with GEANT4 simulation predictions
  using different models.
Only statistical errors are displayed, since most systematic ones cancel.}
\label{fig:ratio12}
\end{figure}

%
\begin{figure}[tbp]
\begin{center}
  \epsfig{figure=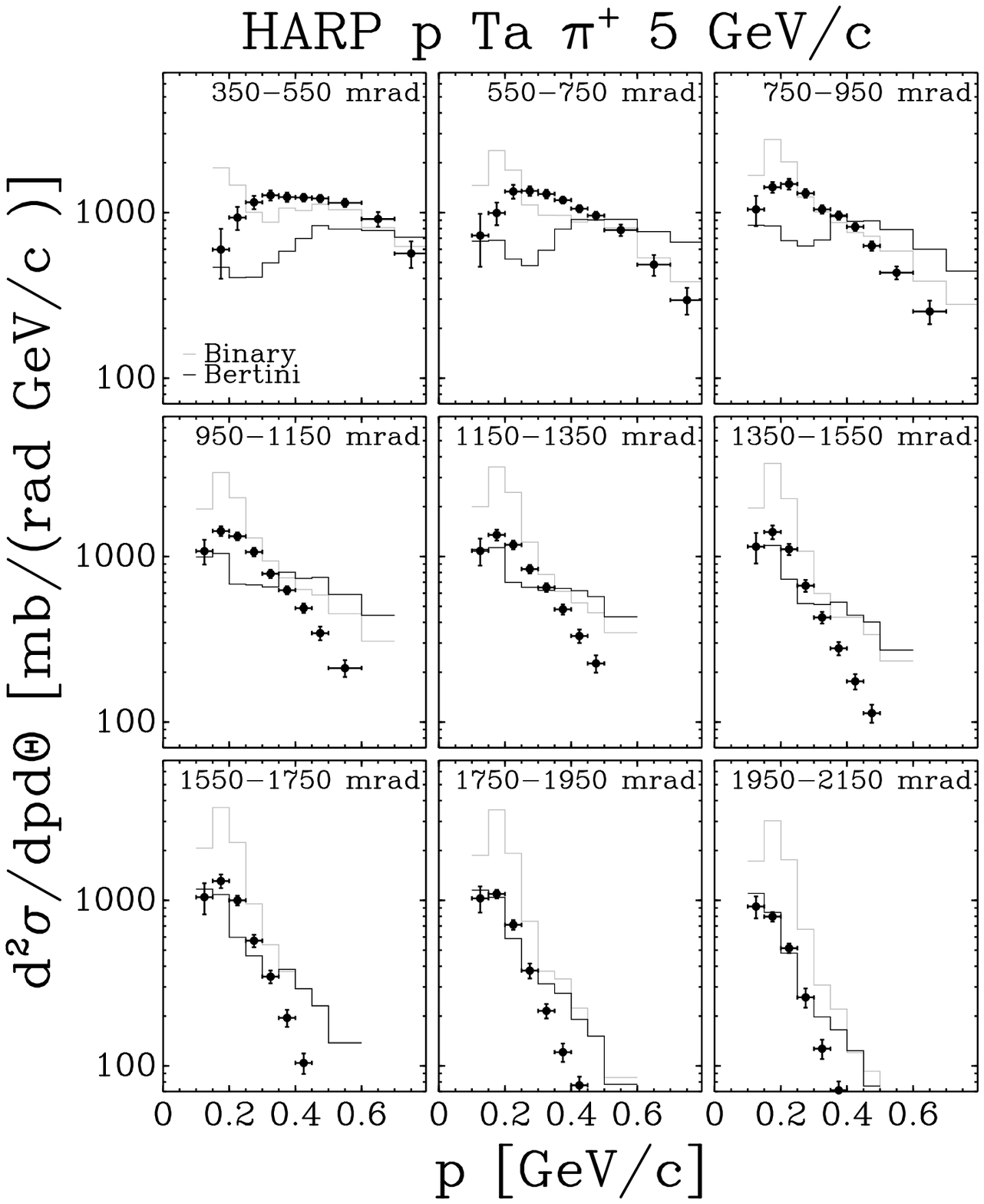,
width=0.48\textwidth}
  \epsfig{figure=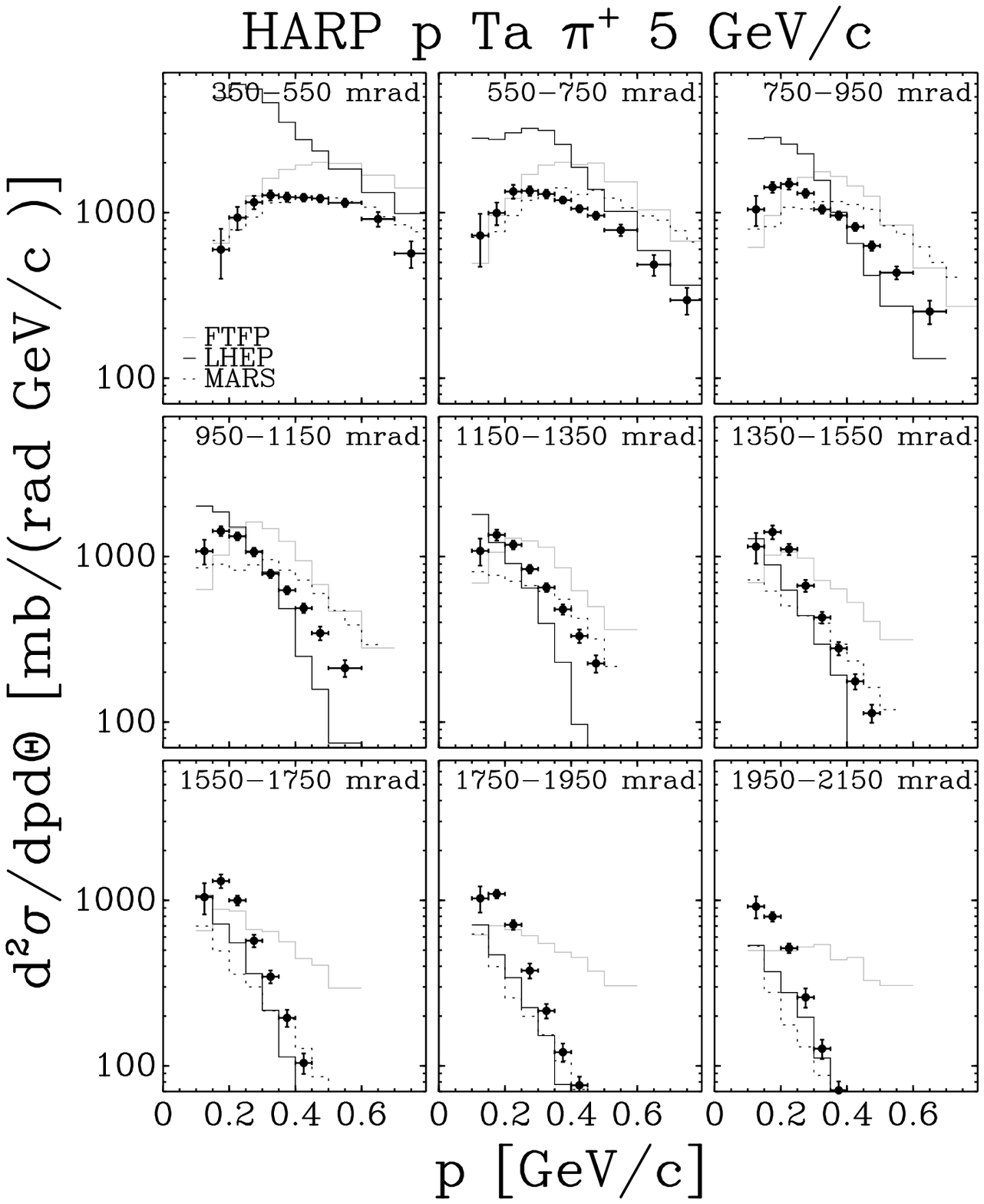,
width=0.48\textwidth}
\end{center}
\caption{
 Double-differential \piplus
 production cross sections for p--Ta at 5~\GeVc and comparison with
 GEANT4 and MARS MC predictions, using several generator models.}
\label{fig:xsTa5GeV}
\end{figure}

\begin{figure}[tbp]
\begin{center}
  \epsfig{figure=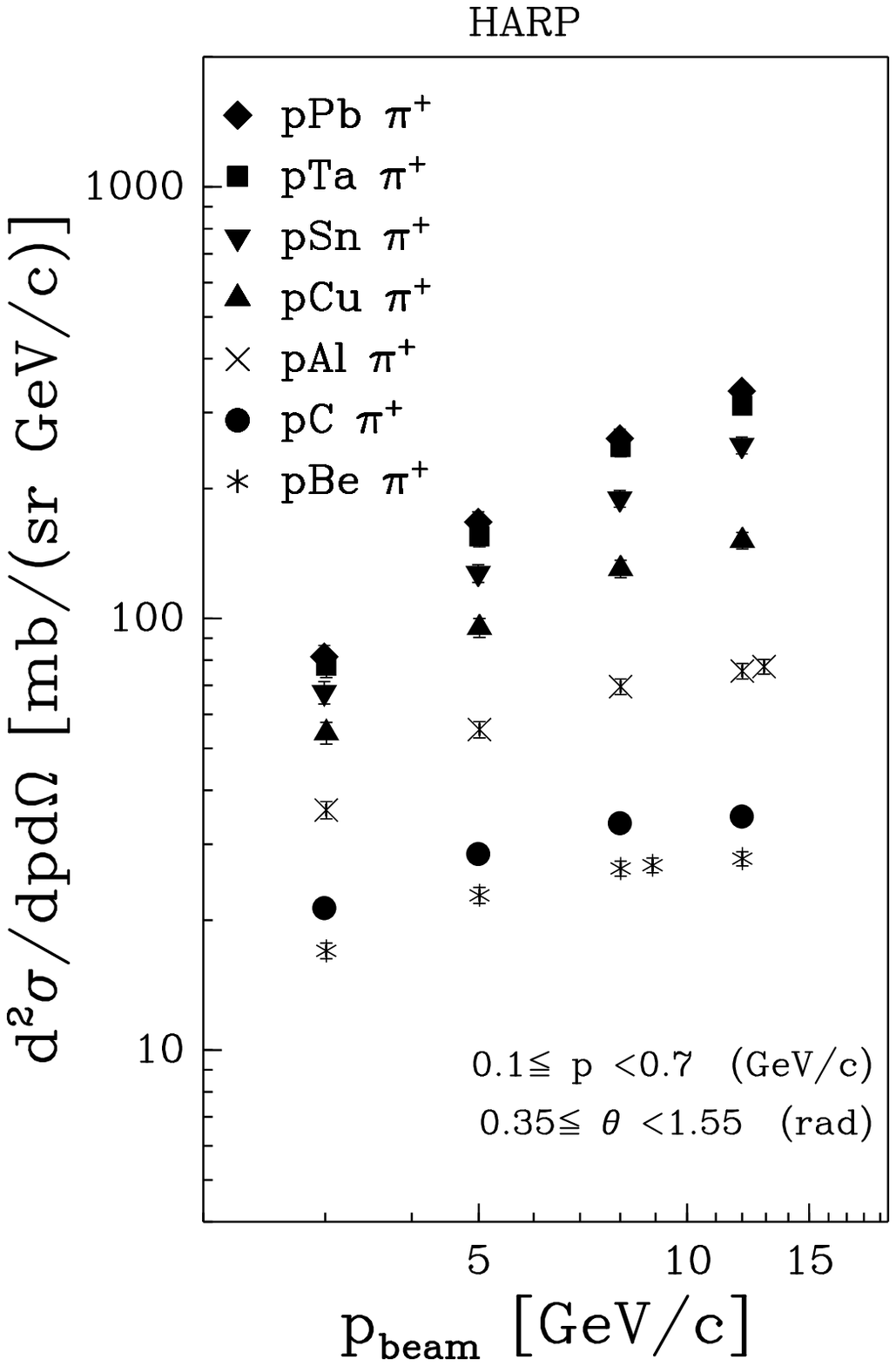,
width=0.38\textwidth, height=0.35\textwidth}
  \epsfig{figure=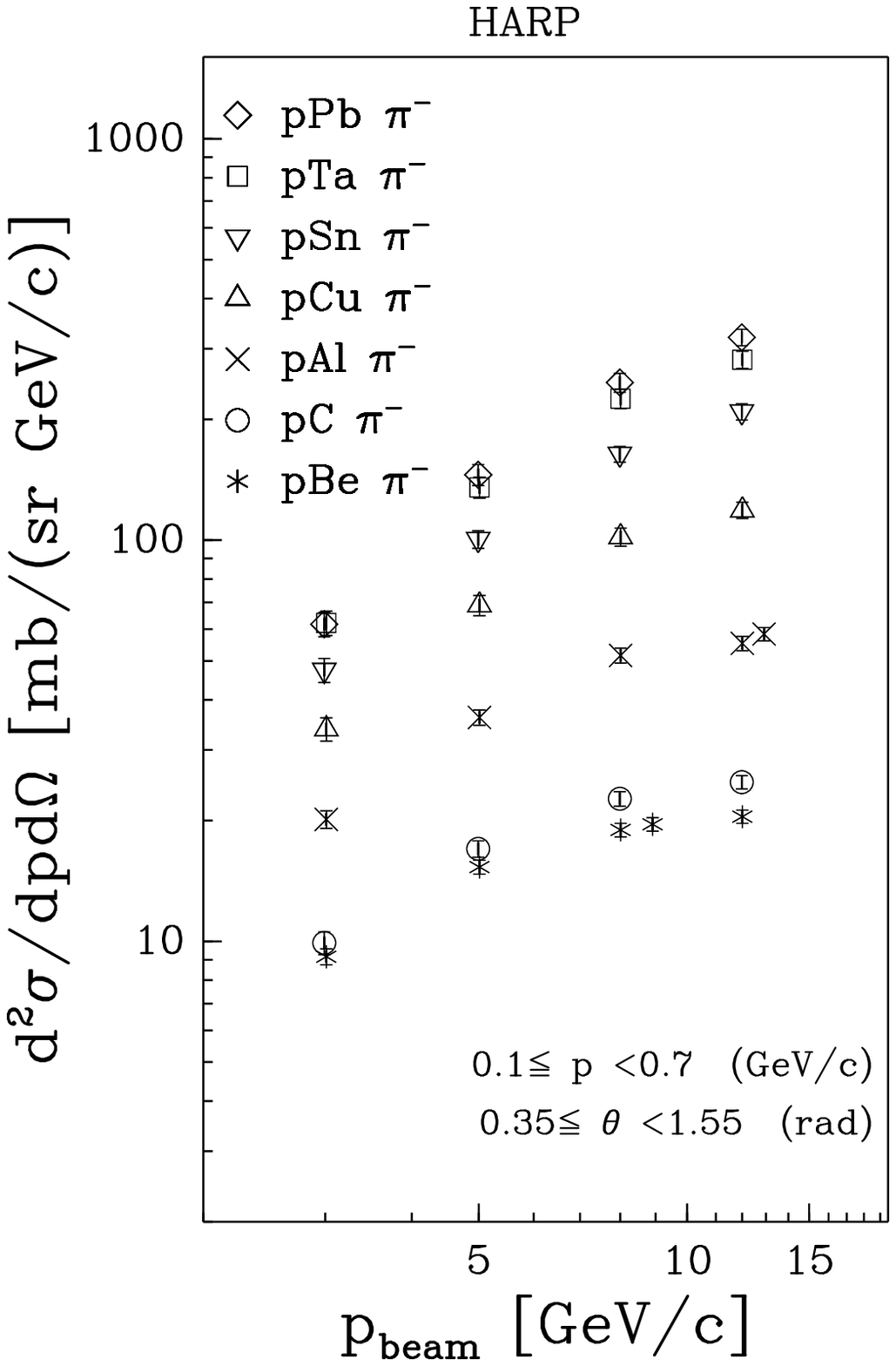,
width=0.38\textwidth, height=0.35\textwidth}
  \epsfig{figure=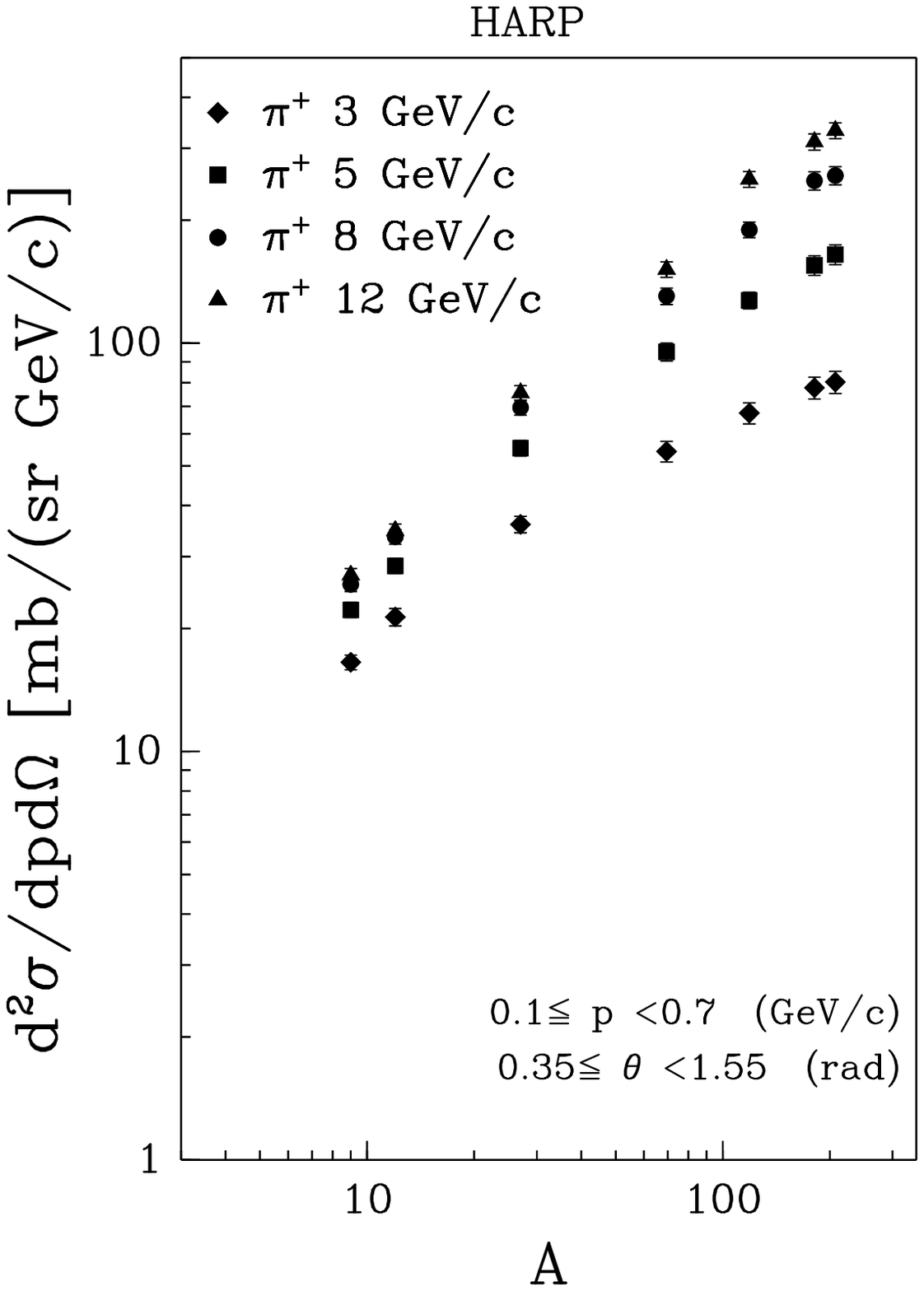,
width=0.38\textwidth, height=0.35\textwidth}
  \epsfig{figure=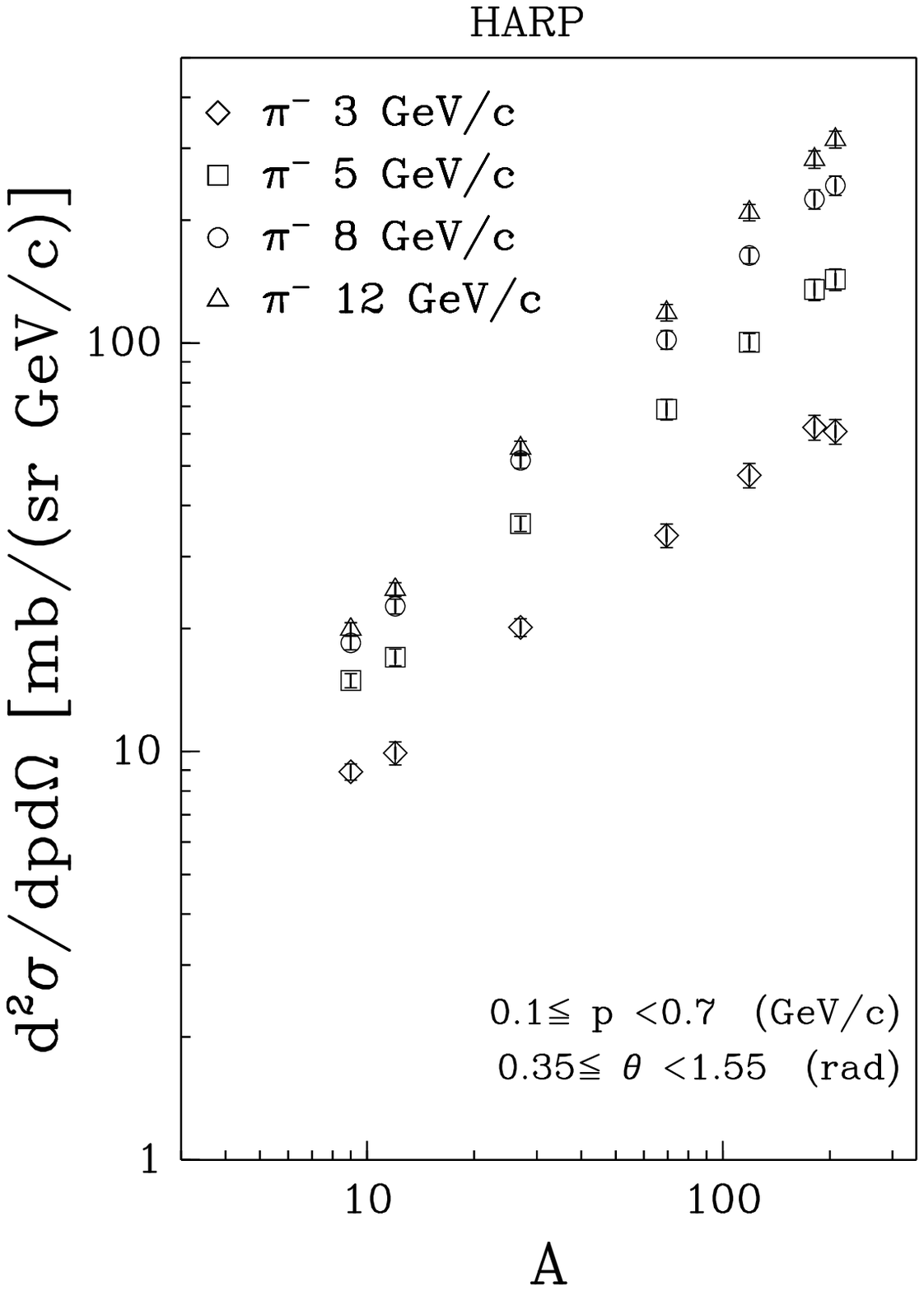,
width=0.38\textwidth, height=0.35\textwidth}
\end{center}
\caption{
 The dependence on the beam momentum and on the atomic number $A$ 
of the \pim (right) and \pip (left) 
  production yields
 in p--Be, p--C, p--Al, p--Cu, p--Sn, p--Ta, p--Pb
 interactions averaged over the forward angular region 
 ($0.350~\rad \leq \theta < 1.550~\rad$) 
 and momentum region $100~\MeVc \leq p < 700~\MeVc$.
 The results are given in arbitrary units, with a consistent scale
 for all panels.
}
\label{fig:xs-a-trend}
\end{figure}

\begin{figure}[tbp]
  \begin{center}
  \epsfig{figure=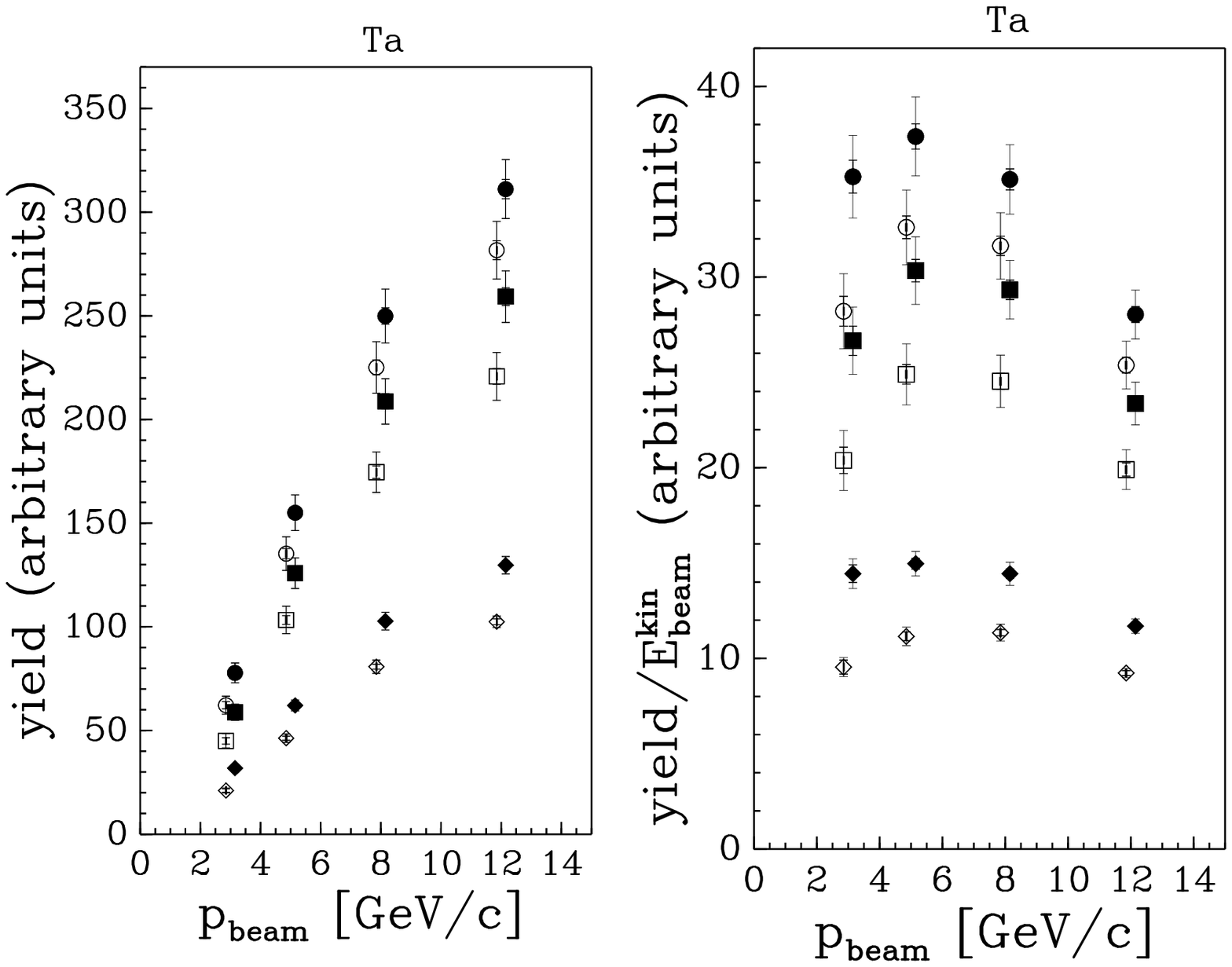,
width=0.48\textwidth}
  \epsfig{figure=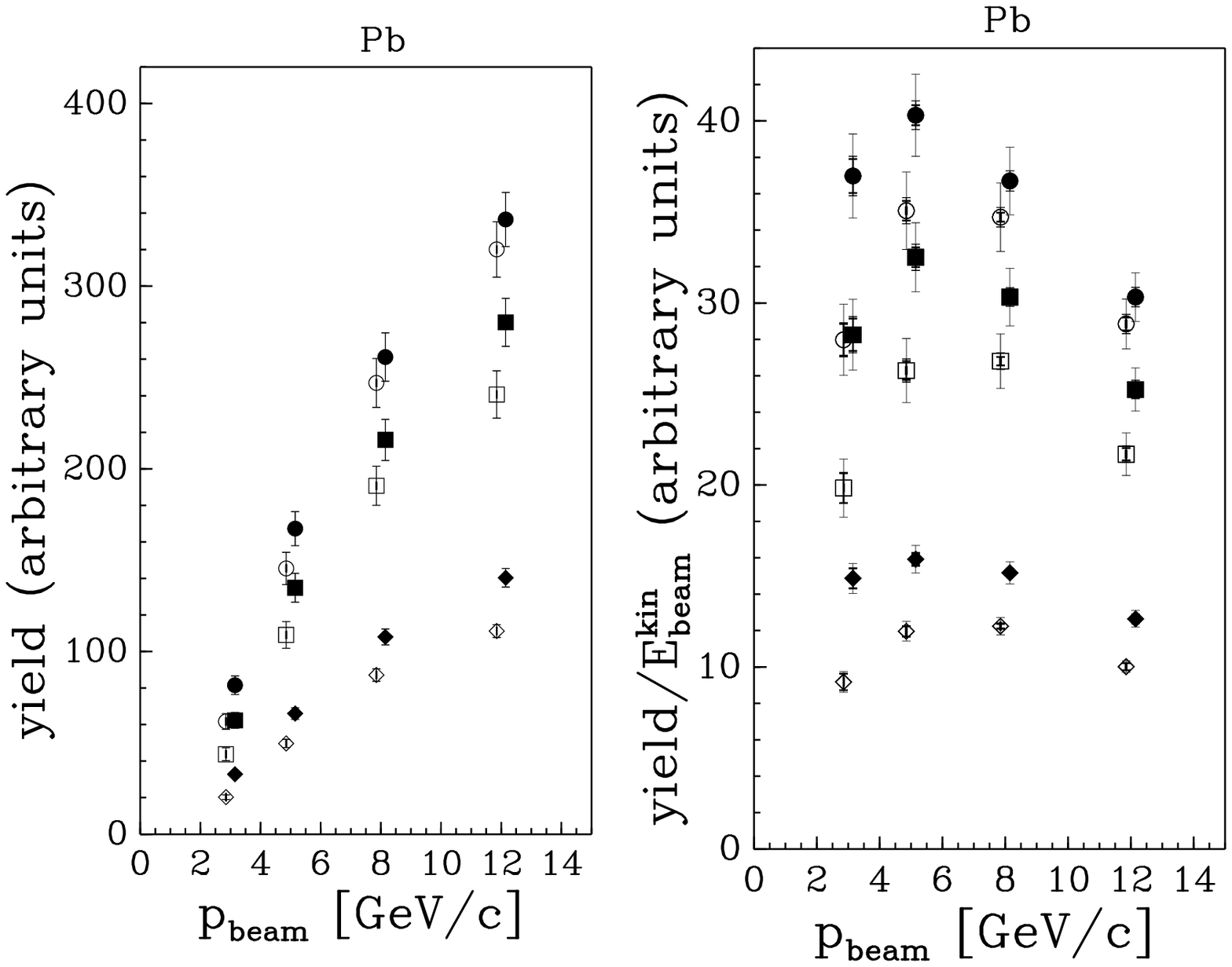,
width=0.48\textwidth}
  \end{center}
\caption{
Predictions of the $\pi^{+}$ (closed symbols) and
$\pi^{-}$ (open symbols) yields for different designs of the neutrino factory
focusing stage.
Integrated yields  and the integrated yields
normalized to the kinetic energy of the proton for 
p--Ta and p--Pb interactions.
The circles indicate the integral over the full HARP acceptance
($100~\MeVc<p<700~\MeVc$ and $0.35 \ \rad <\theta < 1.55 \ \rad$), the
squares are integrated over $0.35 \ \rad <\theta< 0.95 \ \rad$, while the diamonds
are calculated for the smaller angular range and
$250~\MeVc<p<500~\MeVc$.
Although the units are indicated as ``arbitrary'',
for the largest region the yield is expressed as
${{\mathrm{d}^2 \sigma}}/{{\mathrm{d}p\mathrm{d}\Omega }}$ in
mb/(\GeVc~sr).
For the
other regions the same normalization is chosen, but now scaled with the
relative bin size to show visually the correct ratio of number of pions
produced in these kinematic regions.
}
\label{fig:nufact-yield}
\end{figure}

 The experiment
 makes use of a large-acceptance spectrometer consisting of a
 forward and large-angle detection systems.
 A detailed
 description of the experimental apparatus can be found in 
Ref.~\cite{ref:harpTech}.
 The forward spectrometer -- 
 based on large area drift chambers  and a dipole magnet
 complemented by a set of detectors for particle identification~\cite{ref:pidPaper}: 
 a time-of-flight wall, a large Cherenkov detector 
 and an electromagnetic calorimeter  --
 covers polar angles up to 250~mrad which
 is well matched to the angular range of interest for the
 conventional neutrino beams.
 The large-angle spectrometer -- based on a Time Projection Chamber (TPC)
 and Resistive Plate Chambers (RPCs),
 located inside a solenoidal magnet --
 has a large acceptance in the momentum
 and angular range for the pions relevant to the production of the
 muons in a neutrino factory.
 It covers the large majority ($\sim 70\%$) of the pions accepted in the focusing
 system of a typical design.

\vspace*{-0.35cm}
\section{Results obtained with the HARP forward spectrometer}
\vspace*{-0.20cm}

%
The first HARP physics publication~\cite{ref:alPaper} reported measurements of the
$\pi^+$ production cross-section from an aluminum target 
at 12.9~GeV/c proton momentum for the K2K experiment at KEK PS.  
The results  were
subsequently applied to the final neutrino oscillation analysis 
of K2K~\cite{ref:k2kfinal}, allowing a significant reduction 
of the dominant systematic error associated with the calculation of
the so-called far-to-near ratio. 
Our next result~\cite{ref:bePaper} was the measurement of the $\pi^+$ cross-sections 
from a thin 5\%~$\lambda_{\mathrm{I}}$
beryllium target at 8.9~GeV/c proton momentum. 
It contributed to the understanding of 
the MiniBooNE and SciBooNE neutrino fluxes~\cite{ref:MiniBooneRes}. 
They are both produced by the Booster Neutrino Beam at 
Fermilab which originates from protons accelerated to 8.9~GeV/c by 
the booster before being collided against a beryllium target. 

Further, measurements 
of the double-differential production cross-section of $\pi^\pm$ 
in the collision of 12~GeV/c protons and  $\pi^\pm$ 
with thin 5\%~$\lambda_{\mathrm{I}}$ 
carbon target and liquid \nn and \oo targets were performed .
These measurements are important for a precise calculation of 
the atmospheric neutrino
flux and for a prediction of the development of extended air showers.
The results for the pion production on the carbon target, the
ratio \nn/Carbon, and comparison with
 models typically used in air shower simulations~\cite{ref:FWmodels}  
are shown in Figs. \ref{pCpipmModelslog} and \ref{fig:ratio12}~\cite{ref:carbonfw}.
The conclusion of comparing the predictions of the
models to the measured data is that they do predict the ratio 
of cross-sections and often fail in predicting the absolute rates, especially
in certain regions of the phase space.

%
%

In practice production targets are not thin and cascade calculations or
dedicated measurements with 'replica targets' are needed.  HARP has
taken, albeit with somewhat lower statistics, and analyzed p+A
data at different beam momenta with 100\%~$\lambda_{\mathrm{I}}$
targets.  They can be used for parametrizations or tuning of models.
Preliminary spectra are available for 
p + Be, C, Al, Cu, Sn, Ta, Pb interactions at 3 -- 12 GeV/c.
 The measurements are on the tapes and can be
analyzed on demand.

\vspace*{-0.35cm}
\section{Results obtained with the HARP large-angle spectrometer}
\vspace*{-0.20cm}

The HARP TPC is the key detector for the analysis of
tracks emerging from the target at large angles with respect to the
incoming beam direction.
 It suffered from a number of shortcomings that were discovered
 during and after the data taking~\cite{ref:harpTech}.
A description of the measures taken to
correct for the effects of them is 
given in
~\cite{ref:harpTech,ref:harp:tantalum,ref:momscale}.
Wide range of experimental cross-checks
has been employed to assess the momentum scale and momentum resolution
in the HARP TPC, summarized in our recent paper~\cite{ref:momscale}.


A group of people formerly belonging to the HARP collaboration and
subsequently detached themselves from it have been criticizing our 
methods of TPC and RPCs calibration~\cite{DydakCommCal}. Our arguments against this
criticism and for the correctness of our results are presented
in~\cite{OurRebuttals,ref:momscale}.



A first set of results on the production of pions at large angles
have been published by the HARP collaboration in
the papers \cite{ref:harp:tantalum,ref:harp:cacotin},
 based on the analysis of the data in the beginning of
each accelerator spill. Track recognition, momentum determination and particle
identification were all performed based on the measurements made with
the TPC.
The reduction of the data set was necessary to avoid problems in the 
chamber responsible for dynamic distortions to the image of the
particle trajectories as the ion charge was building up during each
spill.
Corrections for such distortions that allow the use of the full statistics
have been developed~\cite{ref:momscale}
and applied in the analysis. The results exploiting the full spill data
have been obtained recently~\cite{FinalProton}. They  are fully 
compatible with the previous ones and cover pion production
by proton beams in a momentum range from
3~GeV/c to  12~GeV/c hitting Be, C, Al, Cu, Sn, Ta and Pb targets with a thickness of
5\%~$\lambda_{\mathrm{I}}$  
in the angular and momentum phase space  
$100~\mbox{MeV/c} \le p < 800~\mbox{MeV/c}$ and 
$0.35~\mbox{rad} \le \theta <2.15~\mbox{rad}$ 
in the laboratory frame.

As an example we show in Fig.~\ref{fig:xsTa5GeV} 
the results for the double-differential cross-sections 
$
{{\mathrm{d}^2 \sigma}}/{{\mathrm{d}p\mathrm{d}\theta }}
$
at  5~GeV/c incident proton beam momentum and Ta target compared to the
respective predictions of several different generator models used in
GEANT4 and MARS simulation packages.
The comparison between data and models 
is reasonable, 
but some discrepancies are evident for some models. 
Discrepancies up to a factor of three are seen. For details see the
full paper~\cite{FinalProton}.
 The dependence on the beam momentum and on the atomic number $A$ of
 integrated yields are presented in Fig.~\ref{fig:xs-a-trend}. 
Predictions of the \piplus and \piminus integrated yields relevant for
the design of the neutrino factory focusing stage 
are given in Fig.~\ref{fig:nufact-yield}.

\vspace*{-0.35cm}
\section{Conclusions}\label{sec:conclusions}
\vspace*{-0.20cm}


The full set of HARP data is in process of publishing now, It  covers
 the pion production by protons and
pions on nuclear targets 
spanning the full periodic table of elements and large solid angle and
 momentum range
in the difficult energy region between
3 and 15 GeV/c of incident momentum.  HARP results fill in an
essential gap in the available experimental information for soft hadron
production and help in the  understanding of
neutrino fluxes in accelerator neutrino experiments, prediction
of atmospheric neutrino fluxes, optimization of a future neutrino
factory design and may be used for improvements of the event
generators for simulation of hadronic interactions.

\vspace*{-0.3cm}
\section*{Acknowledgments}
\vspace*{-0.3cm}
It is a pleasure to thank the organizers and INFN -- Section  Bari for the financial support 
which allowed me to 
participate in the conference and to present 
these results on behalf of the HARP Collaboration.

\end{document}